\documentclass[showpacs,twocolumn,prl,amsmath,amssymb]{revtex4}
\usepackage{graphicx}
\usepackage{dcolumn}
\usepackage{bm}
\begin{document}
\title{
Mott metal-insulator transition in the Hubbard model
}
\author{Fusayoshi J. Ohkawa}
\affiliation{Department of Physics, Faculty of
Science,  Hokkaido University, Sapporo 060-0810, Japan}
\email{fohkawa@phys.sci.hokudai.ac.jp}
\received{25 June 2006; revised manuscript received 1 January 2007}
\begin{abstract} 
The ground state of the Hubbard model is studied within 
the single-site approximation (SSA) and beyond the SSA.
Within the SSA,
the ground state is a typical Mott insulator at the critical point 
$n=1$ and $U/W=+\infty$,
with $n$ being the electron density per unit cell, $W$ the bandwidth of electrons,
and $U$ the on-site repulsion, and  is a normal Fermi liquid   
except for the critical point.
Beyond the SSA,
the normal Fermi liquid is unstable against a non-normal Fermi liquid state
except for a trivial case of $U=0$ such as 
a magnetic or superconducting state in two and higher dimensions.
 In order to explain actual observed metal-insulator transitions,
one or several effects among 
the electron-phonon interaction, 
multi-band or multi-orbital effects, and  effects of disorder
should be considered beyond the Hubbard model.
\end{abstract}
\maketitle

The Mott metal-insulator (M-I) transition is an interesting 
and important issue in solid-state physics \cite{mott}, and a lot of effort 
has been made towards clarifying it \cite{tokura}.
However,  its theoretical treatment is still controversial. 
One of the most contentious issues is whether or not
the transition can be explained within the Hubbard model.

In Hubbard's approximation \cite{Hubbard1,Hubbard2},
a band splits into two subbands when the on-site repulsion $U$ is large enough such that
$U\agt W$, with $W$ the bandwidth $(W>0)$;
the subbands are called the upper Hubbard band (UHB)
and the lower Hubbard band (LHB). 
In Gutzwiller's approximation \cite{Gutzwiller1,Gutzwiller3}, 
a narrow quasiparticle band appears around the chemical potential;
the band and quasiparticles are called the Gutzwiller band 
and quasiparticles. 
It is plausible to speculate that 
the density of states in fact has a three-peak structure,
with the Gutzwiller band between UHB and LHB. 
Both of the approximations are single-site approximations (SSA).
Another SSA theory   \cite{OhkawaSlave} confirms this speculation,
 showing that the Gutzwiller band appears
at the top of LHB when the electron density per unit cell $n$
is less than one, i.e., $n<1$.
According to Kondo-lattice theory 
\cite{Mapping-1,Mapping-2,Mapping-3},
the three-peak structure corresponds to
the Kondo peak between two subpeaks in the Anderson model, 
which is an effective Hamiltonian for the Kondo problem.
An insulating state can only be realized when 
the Fermi surface (FS) of the Gutzwiller quasiparticles vanishes.

\begin{figure}
\centerline{
\includegraphics[width=5.0cm]{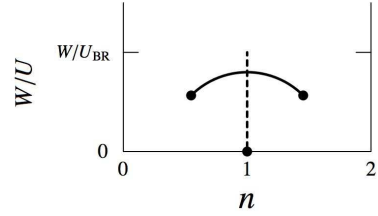}
}
\caption[1]{
Schematic phase diagram for $T=0$~K within the subspace
with no order parameter.   The arc indicates 
a possible but unlikely first-order transition line between metallic states.
Dots indicate critical points.
See the text for the dashed line.
}
\label{fig_phase}
\end{figure}

When $n=1$ and $W/U=0$,
a  localized electron or  a non-interacting spin sits at every unit cell.
In this case, the ground state  is infinitely degenerate and
is a typical Mott insulator.
In Gutzwiller's approximation,  
the ground state is a metal for $n\ne 1$ or $W/U>0$.
According to Brinkman and Rice's theory \cite{brinkman},
when $n=1$ the effective mass of the quasiparticles diverges
at a critical $U \simeq W$, which  is denoted as $U_{\rm BR}$ here.
It may be argued that the FS of the quasiparticles vanishes 
for $n=1$ and $0\le W/U \le W/U_{\rm BR}$, i.e., on the dashed line
in the phase diagram shown in Fig.~\ref{fig_phase}. 
It is curious that no order parameter appears 
in this second-order transition
and every physical property seems to be continuous across the line.
It is probable therefore that  
the critical $U$ is infinite
beyond Gutzwiller's approximation.
One of the purposes of this Letter is to show
that no Mott M-I transition is possible at any finite $U$, 
which contradicts a scenario that actual observed M-I transitions
can be explained within the Hubbard model. 
Hence, the other purpose is to examine
relevant  effects beyond the Hubbard model,
which should be considered to explain the transitions. 

The Hubbard and Anderson models are defined by
\begin{eqnarray}\label{EqHub}
{\cal H} \hspace{-3pt}&=&\hspace{-3pt}
\epsilon_a \sum_{i\sigma}n_{i\sigma}
- \sum_{i\ne j}\sum_{\sigma} t_{ij} a_{i\sigma}^\dag a_{j\sigma}
+ U \sum_{i} n_{i\uparrow} n_{i\downarrow},
\\ \label{EqAnd}
{\cal H}_{A}\hspace{-3pt}&=&\hspace{-3pt}
\sum_{{\bf k}\sigma} E_c({\bf k}) 
c_{{\bf k}\sigma}^\dag c_{{\bf k}\sigma}
+ \! \frac1{\sqrt{N}} \! \sum_{{\bf k}\sigma} \! \left[
V({\bf k})c_{{\bf k}\sigma}^\dag d_\sigma
\!+ (\mbox{h.c.}) \right] 
\nonumber \\ && \quad 
+ \epsilon_d \sum_{\sigma}n_{d\sigma}
+ U n_{d\uparrow} n_{d\downarrow}.
\end{eqnarray}
with $n_{i\sigma}= a_{i\sigma}^\dag a_{i\sigma}$, 
$n_{d\sigma}=d_{\sigma}^\dag d_{\sigma}$, and 
$N$ the number of unit cells. 
When it is assumed there is no order parameter,
the self-energy of electrons for the Hubbard model
is divided into single-site and multi-site ones:
$\Sigma_{\sigma}(i\varepsilon_n, {\bf k})
= \tilde{\Sigma}_{\sigma}(i\varepsilon_n)
+ \Delta \Sigma_{\sigma}(i\varepsilon_n, {\bf k})$.
The single-site one $\tilde{\Sigma}_{\sigma}(i\varepsilon_n)$
is given by that of the Anderson model
when $U$ and the single-site electron lines 
of the Feynman diagrams are the same in  the two models.
The chemical potentials of the two models are denoted by 
$\mu$ and $\tilde{\mu}$, respectively. 
The single-site Green function of the Hubbard model is given by
$R_{\sigma}(i\varepsilon_n) =
(1/N)\sum_{\bf k}1/[\varepsilon_n + \mu - E({\bf k}) - 
\Sigma_{\sigma}(i\varepsilon_n, {\bf k})] $,
with
$E({\bf k}) = \epsilon_a -
(1/N)\sum_{i\ne j} t_{ij} \exp\left[i{\bf k}\cdot
\left({\bf R}_i-{\bf R}_j\right)\right]$.
Here, ${\bf R}_i$ is the position of the $i$th lattice site.
The Green function of the Anderson model is given by
$\tilde{G}_{\sigma}(i\varepsilon_n) =
1/[\displaystyle i\varepsilon_n + \tilde{\mu}
- \epsilon_d - \tilde{\Sigma}_{\sigma}(i\varepsilon_n)
-L_\sigma(i\varepsilon_n)]$,
with
$L_\sigma(i\varepsilon_n) =
(1/N) \sum_{\bf k}
|V({\bf k})|^2 /\left[i\varepsilon_n+\tilde{\mu} - E_c({\bf k}) \right]$.
The condition for the electron lines is 
$R_{\sigma}(i\varepsilon_n) = \tilde{G}_\sigma (i\varepsilon_n)$.
In fact, a set of $\tilde{\mu}-\epsilon_d  = \mu - \epsilon_a $ and
 \begin{equation}\label{EqmappingConditionA}
 L_\sigma(\varepsilon \!+\! i0)=\varepsilon + \mu -\epsilon_a
- \tilde{\Sigma}_\sigma(\varepsilon \!+\! i0) 
-1/R_\sigma(\varepsilon \!+\! i0) ,
\end{equation}
is  a mapping condition to the Anderson model.
First, 
the multi-site self-energy is ignored  
in Eq.~(\ref{EqmappingConditionA}).
The approximation is the best SSA, which
considers all the single-site terms, and is reduced to
a problem of determining and solving the Anderson model  
\cite{Mapping-1,Mapping-2,Mapping-3}.
The SSA is rigorous for infinite dimensions within the Hilbert subspace 
with no order parameter \cite{Metzner}.
The SSA can also be formulated as 
the dynamical mean-field theory (DMFT) \cite{georges,PhyToday} and
the dynamical coherent potential approximation \cite{dpca}.

According to   
Yosida's perturbation theory \cite{yosida} 
and   Wilson's renormalization-group theory \cite{wilsonKG},
provided that the FS of conduction electrons is present,
the ground state of the $s$-$d$ model is a singlet  or 
a normal Fermi liquid (FL). 
Exceptionally, it is a doublet for $J_{s\mbox{-}d}=0$, with
$J_{s\mbox{-}d}$ the $s$-$d$ exchange interaction.
Since the $s$-$d$ model is derived from the Anderson model,
the ground state of the Anderson model is a singlet or a normal FL,
or exceptionally a doublet for 
the just half filling and infinite $U$.
A necessary condition for the presence of  the FS  is 
\begin{equation}\label{EqFL}
\mbox{Im}\left[L_\sigma(\varepsilon+i0) 
\right]_{\varepsilon\rightarrow \pm 0} < 0. 
\end{equation}
This condition is called the FS condition in this Letter.

The density of states for $U=0$ is given by
$\rho_0(\varepsilon) =(1/N)\! \sum_{\bf k}\!
\delta[\varepsilon \!-\! E({\bf k})]$.
It is assumed that the FS is present for $U=0$ or that
$0<n<2$ and there is no gap in $\rho_0(\varepsilon)$.  

As a preliminary, 
a Lorentzian model  is examined, where
$\rho_0(\varepsilon)$ is given by $\rho_L(\varepsilon)\equiv$
$(1/\pi)\Delta/[(\varepsilon-\epsilon_a)^2 + \Delta^2]$,
with $\Delta>0$. 
According to Eq.~(\ref{EqmappingConditionA}),  
$L_\sigma(\varepsilon\!+\!i0) = - i\Delta$ \cite{georges}.
Since the FS condition (\ref{EqFL}) is satisfied for the Anderson model,
the ground state of the Hubbard model 
is a normal FL except for $n=1$ and $W/U=0$.

In order to examine a non-Lorentzian model of $\rho_n(\varepsilon)$,
the following model is first examined:
\begin{equation}
\rho_0(\varepsilon) = - \frac1{\pi} \mbox{Im}
\int d\varepsilon^\prime \frac{\rho_n(\varepsilon^\prime)}
{\varepsilon-\varepsilon^\prime + i\delta\Delta},
\end{equation}
with $\delta>0$.
According to Eq.~(\ref{EqmappingConditionA}) and 
 the inequality
$\int d\varepsilon^\prime \rho_0(\varepsilon^\prime)
[S_1(\varepsilon,\varepsilon^\prime) +x]^2/
[S_1^2\varepsilon,\varepsilon^\prime) +S_2^2(\varepsilon)
] >0$ for any real $x$, with 
$S_1(\varepsilon,\varepsilon^\prime)$
and $S_2(\varepsilon)$ being real functions defined by
$S_1(\varepsilon,\varepsilon^\prime) +i S_2(\varepsilon)=
\varepsilon +\mu - \epsilon_a - \varepsilon^\prime 
- \tilde{\Sigma}_\sigma(\varepsilon+i0)$,
it follows that
$\mbox{Im}L_\sigma(\varepsilon+i0) \le 0$.
The equality is only possible provided that $\delta=+0$. 
The FS condition (\ref{EqFL}) is satisfied for $\delta>0$.
The ground state for $\rho_n(\varepsilon)$ is obtained
in the adiabatic process $\delta \rightarrow+0$. 
This gives a normal FL except for $n=1$ and $W/U=0$, 
which may be degenerate with a non-normal FL provided that
$\mbox{Im}[L_\sigma(\varepsilon+i0)
]_{\varepsilon\rightarrow \pm 0,\delta\rightarrow +0}=0$.
No symmetry breaking occurs in the adiabatic process.

Since symmetry breaking is caused by 
Weiss mean fields, which are intersite effects,
symmetry breaking is definitely impossible in the SSA, as is shown above.
Under the SSA, therefore, the adiabatic continuation \cite{AndersonText} 
as a function of $U$ holds.
The ground state for the non-Lorentzian $\rho_n(\varepsilon)$ 
is the normal FL
obtained in the adiabatic process. 
The self-energy is expanded as
%
$\tilde{\Sigma}_\sigma(\varepsilon  +  i0) =
\tilde{\Sigma}_\sigma(0)
+ \bigl(1  -  \tilde{\phi}_\gamma \bigr)\varepsilon 
+ \bigl(1  - \tilde{\phi}_s \bigr)  \frac1{2}\sigma g \mu_BH 
 + O(\varepsilon^2)$
at $T=0$~K
in the presence of an infinitesimally small Zeeman energy $g\mu_BH$,
with $\tilde{\Sigma}_\sigma(0)$, $\tilde{\phi}_\gamma$, 
and $\tilde{\phi}_s$ all  being real.
According to the FS sum rule 
\cite{Luttinger2}     
the electron density $n$ is given by
$n = (1/N) \sum_{{\bf k}\sigma}
\theta\bigl( [ \mu - E({\bf k}) - \tilde{\Sigma}_\sigma(0)
]/W \bigr)$,
with $\theta(x<0)=0$ and $\theta(x>0)=1$.
The density of states  
is given by
$\rho(\varepsilon) =- (1/\pi)\mbox{Im} R_\sigma(\varepsilon+i0)$.
When $n$ is kept constant,
$\mu - \tilde{\Sigma}_\sigma(0)$,
$\rho(0)$, and $L_{\sigma}(+i0)$ 
do not depend on $U$.
Since $\mbox{Im}L_\sigma(+i0)<0$ for  $U=0$, 
$\mbox{Im}L_\sigma(+i0)<0$ for $U/W\ge 0$.
When $\rho_n(\varepsilon)$ is continuous and finite,
$L_\sigma(\varepsilon+i0)$ is continuous
and the FS condition  is satisfied so that
the FL obtained in the adiabatic process
is not degenerate with a non-normal FL.
Provided that $\rho_n(\varepsilon)$ is discontinuous or divergent,
degeneracy is possible for certain $n$ corresponding to
the discontinuity or divergence of $\rho_n(\varepsilon)$
according to the FS sum rule.

In Gutzwiller's approximation,    
$\tilde{\phi}_\gamma \propto 1/|1-n|$ for $W/U=0$.
Since $\tilde{\phi}_\gamma$ diverges as $n\rightarrow1$, 
$\mbox{Re}[\tilde{\Sigma}_\sigma(\varepsilon \!+\! i0)
]_{n\rightarrow1}$
is at least discontinuous at $\varepsilon=0$ so that
$\mbox{Im}[\tilde{\Sigma}_{\sigma}(\epsilon +i0)
]_{\epsilon\rightarrow \pm0,n\rightarrow1}$ diverges logarithmically 
according to the Kramers-Kronig relation. 
Then, 
$\rho(\varepsilon\rightarrow \pm 0)=0$ for $n\rightarrow1$ and $W/U=0$, 
although $\rho(0)>0$ for $n=1$ and  $W/U\rightarrow0$. 
%
There is a discontinuity in $\rho(\varepsilon)$
as a function of $\varepsilon$, $n$, and $W/U$.
The critical point of $n=1$ and $W/U = 0$
is unconventional within the SSA.

If  the mapping to the Anderson model is not unique
or physical quantities such as $\tilde{\phi}_\gamma$ and
$\tilde{\phi}_s$ are multi-valued functions of $U$,
a first-order transition between metallic states occurs. 
The adiabatic continuation still holds, for example, along
a route around one of the critical points at the ends of
the first-order transition line.
Since the FS sum rule, 
$\rho(0)$,  and $L_\sigma(+i0)$ are all the same 
in the two metallic states on different sides of the line, 
the occurrence of the first-order transition is unlikely. 
The transition never occurs in the Lorentzian model because the mapping is unique.
The transition is shown 
on a schematic phase diagram in Fig.~\ref{fig_phase}.

The occurrence of a first-order M-I transition is suggested 
by numerical theories using the Monte Carlo method
\cite{imada} and DMFT \cite{kotliar}.
In these theories, the compressibility 
$\tilde{\chi}_c(0) = dn(\mu)/d\mu$ shows a rapid change
\cite{ComComp}.
When the rapid change is really a jump,
the phase diagram for $T>0$~K is like that shown in Fig.~1 of 
Ref.~\onlinecite{kotliar}.
The phase diagram suggests that
the first-order M-I transition occurs even at $T=0$~K.
If this is the case
within the Hilbert subspace with no order parameter, however, it cannot 
be an M-I transition, but rather it must be a transition between metallic states.

The irreducible spin polarization function 
is also divided into single-site and multi-site ones:
$\pi_s(i\omega_l,{\bf q}) =
\tilde{\pi}_s(i\omega_l) +\Delta\pi_s(i\omega_l,{\bf q}) $.
The single-site one $\tilde{\pi}_s(i\omega_l)$ is given by
that of the Anderson model.
The spin susceptibilities of the Anderson and Hubbard models are given, 
respectively, by
$\tilde{\chi}_s(i\omega_l) =
2\tilde{\pi}_s(i\omega_l)/[
1 - U \tilde{\pi}_s(i\omega_l)]$
and
$\chi_s(i\omega_l,{\bf q}) =
2\pi_s(i\omega_l,{\bf q}) /[
1 - U \pi_s(i\omega_l,{\bf q})]$.
In Kondo-lattice theory,
an intersite exchange interaction 
$I_s(i\omega_l,{\bf q})$ is defined by
$\chi_s(i\omega_l,{\bf q}) =
\tilde{\chi}_s(i\omega_l)/[
1 - (1/4)I_s(i\omega_l,{\bf q}) \tilde{\chi}_s(i\omega_l)]$.
When $U/W\agt 1$,   
$I_s(i\omega_l,{\bf q}) \simeq 2 U \Delta\pi_s(i\omega_l,{\bf q})$,
where $O[1/U\tilde{\chi}_s(i\omega_l)]$ terms are ignored.
The exchange interaction is composed of three terms
\cite{exchange}.
The first is the superexchange interaction.
According to field theory, 
it arises from the exchange of a pair excitation of electrons between
LHB and UHB.
When it is assumed the widths of LHB and UHB are vanishingly small,
the strength of the interaction between nearest neighbors is $J = - 4|t|^2/U$,
with $t$ the transfer integral between nearest neighbors 
\cite{ComJ}.
The second is an exchange interaction
arising from the exchange of a pair excitation
of the quasiparticles.
The strength of the interaction is proportional to the bandwidth of the quasiparticles.
It is antiferromagnetic when the nesting of the FS is sharp or
the chemical potential lies around the band center of the quasiparticles,
and it is ferromagnetic when the chemical potential lies
around one of the band edges of the quasiparticles.
The third term corresponds to the mode-mode
coupling term in the self-consistent renormalization  theory
of spin fluctuations  \cite{moriya}, which
is relevant for $U/W\alt 1$. 
According to the Ward relation \cite{ward},
the static component of the single-site
irreducible vertex function in spin channels
is given by $\tilde{\lambda}_s = \tilde{\phi}_s[1 -U \tilde{\pi}_s(0)]$.
When this is used, 
the mutual interaction between the quasiparticles is given by
$(U \tilde{\lambda}_s)^2
[\chi_s(i\omega_l,{\bf q}) - \tilde{\chi}_s(i\omega_l)]=
\tilde{\phi}_s^2 I_s^*(i\omega_l,{\bf q})$, with
$I_s^*(i\omega_l,{\bf q})=I_s(i\omega_l,{\bf q})
/[1 - (1/4)I_s(i\omega_l,{\bf q}) \tilde{\chi}_s(i\omega_l)]$;
the single-site term is subtracted because it is considered 
in the SSA. The mutual interaction mediated by spin fluctuations
is essentially the same as that due to the exchange interaction
$I_s(i\omega_l,{\bf q})$ or $I_s^*(i\omega_l,{\bf q})$.
In Kondo-lattice theory, an unperturbed state
is constructed in the non-perturbative SSA theory \cite{ComSSA} and 
intersite effects are  perturbatively considered 
in terms of $I_s(i\omega_l,{\bf q})$ 
\cite{comD}.
 
When the Fock-type self-energy due to 
the superexchange interaction $J$  is considered, 
the bandwidth of the quasiparticles is given by
$W^* \propto (W/\tilde{\phi}_\gamma) + c_{J} |J| $,
with $c_J =O(1)$ \cite{phase-diagram}.
When this renormalization is considered,
the critical point of $n=1$ and $W/U=0$ turns conventional;
there is no discontinuity in $\rho(\varepsilon)$  
as a function of $\varepsilon$, $n$, and $W/U$.

An order parameter can appear in two dimensions or higher. 
The stability of a normal FL  
against an ordered state with an order parameter can be 
examined when the corresponding response function 
is perturbatively considered in terms of
$I_s(i\omega_l,{\bf q})$ or $I_s^*(i\omega_l,{\bf q})$.
When $I_s(i\omega_l,{\bf q})$ is strong,
for example, the FL is unstable against a magnetic state.
The N\'{e}el temperature $T_N$ is defined as 
the highest value of $T_N$ determined by 
$[\chi_s(0,{\bf q})]_{T=T_N} \rightarrow +\infty$
as a function of ${\bf q}$.
When $I_s(i\omega_l,{\bf q})$ is so weak that 
$[\chi_s(0,{\bf q})]_{T=0\hspace{1pt}{\rm K}} < +\infty$ 
for any ${\bf q}$,
the FL is stable against any magnetic state.

The energy scale of local quantum spin fluctuations is defined by 
$k_BT_K=1/[\tilde{\chi}_s(0)]_{T=0\hspace{1pt}{\rm K}}$,
where $T_K$ is the so-called Kondo temperature.
In accordance with the $T$-dependent crossover between 
a localized spin for $T\gg T_K$ and a normal FL for $T\ll T_K$
in the Kondo problem \cite{wilsonKG},
magnetism for $T_N \gg T_K$ is characterized as local-moment magnetism 
and magnetism for $T_N \ll T_K$ is characterized as itinerant-electron magnetism
 \cite{phase-diagram}.
 The magnetism crossover is simply 
 a Mott M-I crossover.

When $I_s^*(i\omega_l,{\bf q})$ is weak or strong, 
the FL is unstable against an anisotropic superconducting (SC) state 
at least at $T=0$~K,  provided that no disorder exists.
When $I_s^*(i\omega_l,{\bf q})$ is antiferromagnetic,
the FL is unstable against a singlet SC state.
For example, it has been proposed \cite{highTc1,highTc2} 
that the condensation of $d\gamma$-wave Cooper pairs of
the Gutzwiller quasiparticles due to the superexchange interaction
is responsible for high-$T_c$ superconductivity
\cite{bednortz}, which occurs in the vicinity of the Mott M-I crossover.
When $I_s^*(i\omega_l,{\bf q})$ is ferromagnetic,
the FL is unstable against a triplet SC state.

The FL  can also be unstable against 
a bond-order (BO) state, whose order parameter
is $\bigl<a_{i\sigma}^\dag a_{j\sigma^\prime}\bigr>_{i\ne j}$,
and a flux state, which is a superposition of
several BO states.
Within Kondo-lattice theory, magnetic or SC states are more stable 
than BO and flux states.

When $U/W\alt 1$, 
the conventional perturbation in terms of $U$ is more useful than that
in terms of $I_s(i\omega_l,{\bf q})$. 
A similar argument to that for $U/W\agt 1$ applies in this case.
When the nesting of the FS is sharp,
a non-interacting electron gas is unstable against a spin density wave.
When an interaction between
electrons given by $U^2\chi_s(i\omega_l,{\bf q})$ is considered,
the electron gas is unstable against an anisotropic SC state 
at least at $T=0~$K, provided that no disorder exists.

No order parameter appears in one dimension.
The FL  for $U/W\agt1$ 
becomes a Tomonaga-Luttinger  liquid  when $I_s(i\omega_l,{\bf q})$ is 
perturbatively treated, as
the electron gas does when  $U$ is perturbatively treated.
It is probable that Lieb and Wu's insulating state 
for  the just half filling \cite{Lieb-Wu} can only be obtained by 
non-perturbative theory. 

In this analysis,
the ground state  is not a paramagnetic Mott insulator
except for $n=1$ and $W/U=0$. 
However, since the nature of actual M-I transitions
cannot be explained within the Hubbard model,
various effects should be considered 
in a multi-band or multi-orbital model.
Changes of  lattice symmetries or jumps in
lattice constants are often observed  \cite{tokura},
which implies that 
the electron-phonon interaction should also be considered. 
A relevant electron-phonon interaction must arise from
spin and orbital channels \cite{el-ph}.
Cooperative Jahn-Teller or orbital
ordering must be responsible for the change of lattice symmetries.
As well as the electron-phonon interaction,
the orbital-channel exchange interaction  can play a role  
in the ordering. 

The FS sum rule holds for the Gutzwiller quasiparticles.
The ordinary rule holds in the absence of an order parameter, and a modified 
rule holds even when the Brillouin zone is folded by an antiferromagnetic or 
orbital order parameter.
Then, Wilson's classification  
of crystalline solids into metals and insulators \cite{wilson}
can apply to M-I transitions.
Two types of M-I transitions are possible according to 
the band structures of the quasiparticles
in the absence and presence of an order parameter: 
between a metal and an insulator 
and between a compensated metal and an insulator.
The Kondo temperature $T_K$ can be different in
metallic and insulating phases of a first-order M-I transition. 
 In the metallic phase, $T_K$ is higher than $T$
 and the quasiparticles are well defined.
If $T_K$  is lower than $T$ in the insulating phase,
the quasiparticles are not well defined.
In such a case, the M-I transition is a transition
between a high-$T_K$ itinerant-electron phase
and a low-$T_K$ local-moment phase.
Change of lattice symmetries or jumps in
lattice constants must play a crucial role in
any first-order M-I transition.

Since disorder, either small or large, must always exist,
Anderson localization and other effects due to disorder
\cite{phase-diagram} can play a role in actual M-I transitions or crossovers. 

In conclusion,
the ground state of the Hubbard model is not a paramagnetic
Mott insulator except for $n=1$ and $W/U=0$ in two and higher dimensions.
In order to explain actual M-I transitions,
the electron-phonon interaction, 
multi-band or multi-orbital effects, and the effects of disorder
should be considered beyond the Hubbard model.
Except for the low-$T_K$ local-moment phase of electrons,
whether a crystalline solid is a metal or an insulator 
can be explained by an extended Wilson's classification
of the band structure of quasiparticles in the absence or presence of 
an order parameter.


\begin{thebibliography}{}
\bibitem{ComComp}
The long range Coulomb interaction
exists in actual crystalline solids. Since it requires the charge neutrality,
the electron density must be kept constant. 
Therefore, $dn(\mu)/d\mu=0$ identically.
The compressibility is never any relevant
property for actual M-I transitions.
\bibitem{ComJ}
When nonzero bandwidths of LHB and UHB are considered, 
$|J|$ becomes substantially smaller than  $4|t|^2/U$.
%
In a multi-band model with a strong enough Hund coupling,
the superexchange interaction can be ferromagnetic, as is shown in 
Ref.~\onlinecite{exchange}.
\bibitem{ComSSA} 
When the multi-site self-energy is considered in the mapping condition
(\ref{EqmappingConditionA}),
the single-site and multi-site terms depend on each other.
In principle, therefore,
they should be self-consistently calculated with each other.
\bibitem{comD} 
Kondo-lattice theory can also be formulated as $1/d$ expansion theory,
with $d$ being the spatial dimensionality. 
Weiss mean fields of magnetism are of leading order in $1/d$
as well as local spin fluctuations. Those of 
isotropic $s$-wave superconductivity and CDW,
which are relevant for  $U/W<0$,
are also  of leading order in $1/d$ 
as well as the corresponding local fluctuations. 
In any single-band model,
all the other effects are  of higher order  in $1/d$.
\end{thebibliography}
\end{document}